# Arsenite sorption and co-precipitation with calcite

G. Román-Ross[1#], G. J. Cuello[2], X. Turrillas[3] A. Fernández-Martínez[2] and L. Charlet[4]

[1] Dept. Of Chemistry, Faculty of Sciences, University of Girona, Campus de Montilivi, 17071 Girona, Spain

[2] Institut Laue-Langevin (ILL), 6, Rue Jules Horowitz, B.P.156, 38042 Grenoble Cedex 9, France

[3] Eduardo Torroja Institute for Construction Sciences, CSIC, C/. Serrano Galvache s/n 28033 Madrid Spain

[4] Environmental Geochemistry Group, LGIT-IRIGM, University of Grenoble I, B. P. 53, 38041 Grenoble Cedex 9, France

[#] Author to whom correspondence should be addressed. Present address: Dept. Of Chemistry, Faculty of Sciences, University of Girona, Campus de Montilivi, 17071 Girona, Spain. E-mail: gabriela.roman@udg.es

**Abstract**

Sorption of As(III) by calcite was investigated as a function of As(III) concentration, time and pH. The sorption isotherm, *i.e.* the log $\Gamma_{As(III)}$ *vs.* log $[As(OH)_3°/ As_{sat}]$ plot is S–shaped and has been modelled on an extended version of the surface precipitation model (Farley et al., 1985; Wersin et al., 1989). At low concentrations, $As(OH)_3°$ is adsorbed by complexation to surface Ca surface sites, as previously described by the X-ray standing wave technique (Cheng et al., 1999). The inflexion point of the isotherm, where $As(OH)_3°$ is limited by the amount of surface sites ($S_T$), yields 6 sites $nm^{-2}$ in good agreement with crystallographic data. Beyond this value, the amount of sorbed arsenic increases linearly with solution concentration, up to the saturation of arsenic with respect to the precipitation of $CaHAsO_{3(s)}$. The solid solutions formed in this concentration range were examined by X-ray and neutron diffraction. The doped calcite lattice parameters increase with arsenic content while *c/a* ratio remains constant. Our results made on bulk calcite on the atomic displacement of As atoms along [0001] direction extend those published by Cheng et al., (1999) on calcite surface. This study provides a molecular-level explanation for why As(III) is trapped by calcite in industrial treatments.

# 1. Introduction

Adsorption and co-precipitation have been widely described as two active mechanisms to trap contaminants from natural and polluted waters (Sposito, 1984; Langmuir, 1997; Stumm and Morgan, 1996). Uptake of contaminants in solid phases can remove ions from solution retarding their transport. When a contaminant is incorporated in the bulk rather than simply adsorbed at the surface, it is less available and can be considered 'immobilized' in the environment at least until the host phase is dissolved. Solid solution formation involving calcite is, therefore, likely to play an important role in retarding the migration of contaminants in natural environments where calcite can precipitate as fine suspension of particles, fine fracture fillings, etc. (Davis et al., 1987). Incorporation of divalent cations in calcite has been widely described (Comans and Middleburg, 1987; Zachara et al., 1991; Reeder, 1996; Tesoreiro and Pakow, 1996; Pokrovsky et al., 2000 and references therein) but much less is known about the incorporation of trivalent elements (Cheng et al., 1999; Stipp et al., 2003). Heterovalent solid solutions are by far more complex than liked-charged ion substitutions, as it precludes simple isomorphic substitution (Curti et al., 2005). Arsenite incorporation in calcite bulk phase is potentially important since many arsenic affected waterbodies are at equilibrium or slightly supersaturated with respect to calcite. This is the case in the Bangladesh-Bramapoutre delta upper aquifers ((Kinniburgh and Smedley, 2001), in the Argentinean Pampa aquifers (Smedley et al., 2002), in eutrophic lakes (Ramisch et al., 1999) as well as in contaminated industrial sites (Smedley and Kinniburgh, 2002; Plant et al., 2005; Juillot et al., 1999).

In aquifers and soils, arsenic is present mainly under two oxidation states, *i.e.* as arsenite (III) and arsenate (V) inorganic hydro-oxylated complex, respectively. In moderately oxidizing to moderately reducing solutions, the species $H_3AsO_3^o$ or one of its deprotoned equivalents is considered to be the dominant form of dissolved arsenic (Wood et al., 2002; Pokrovski et al., 1996). In water at 25ºC, As(III) is only present in the form of arsenous acid $H_3AsO_3^o$, also denoted $As(OH)_3^o$, up to pH 8 ($pKa_1$ = 9.32; (Pokrovski et al., 1996)), whereas various species of the arsenate (V) acid occur depending on pH (the first, second and third $H_3AsO_4$ pKa values are 2.24, 6.94 and 12.19, respectively) (Sadiq, 1995).

Arsenic oxyanions chemisorption on Fe and Al oxide/hydroxides solid colloid surface is believed to be a common mechanisms leading to As coprecipitation in soils and sediments and has been widely studied (see review by Plant et al., 2005). Sorption on carbonates has been much less investigated (Cheng et al., 1999; Roman-Ross et al., 2003). Arsenic oxyanions can gradually concentrate on colloid surfaces to a level high enough to precipitate as discrete, or mixed, As solid phases. After the exhaustion of reactive Fe, dominant forms are controlled by the reactive levels of Ca in calcareous soils. Thus, the solubility of these minerals can control the arsenic concentration in the water, but due to the high solubility of *e.g.* calcium arsenate, this mechanism does not allow to limit arsenic concentration to acceptable levels, *e.g.* in contaminated sites. Indeed, over a long period of time, arsenic can be mobilized due to the high solubility of calcium arsenites and arsenates (Swash et al., 2000) and to consumption of lime high pH buffer capacity in sites where lime was applied.

Lime (CaO) is by far the most common method of treating As-rich industrial waste containing As (Bothe and Brown, 1999). As immobilization and formation of Ca–As precipitates have been widely studied (Dutré and Vandercasteele, 1995; Dutré and Vandercasteele, 1998; Dutré et al., 1999 and Vandercasteele et al., 2002). These studies have shown that the precipitation of $Ca_3(AsO_4)_2$ and $CaHAsO_3$ controls the immobilization of As in contaminated natural environments, which have been treated with lime among other raw materials. It is well known that $CaHAsO_3$ is very stable under the high pH (12-13) values provided by lime. However, over longer time periods, the lime progressively reacts with carboanions and the carbonation of lime leads to calcium carbonate precipitation. During this process the CaHAsO$_3$ solid phase is dissolved as pH decreases and, part of the arsenite may be trapped inside the new calcite crystal structure.

The main objective of this work is to study As(III) adsorption and co-precipitation processes in presence of calcite to establish the conditions under which As(III)-calcite surface complexes and solid solutions can be formed. Furthermore, it is investigated the range of As(III) substitution that calcite can include in its crystal structure.

## 2. Materials and experimental methods

All chemicals were reagent grade and used without further purification. Solutions were prepared with Milli-Q(18 MΩ-cm) water. As(V) stock solution was prepared from reagent grade As(III) stock from NaAsO$_2$ (J. T. Baker).

Sorption experiments were conducted using a natural calcium carbonate, Mikhart® 130 from Provençale S.A. (Cases de Pene, France). The crystallography of calcite has been confirmed with X-ray diffraction. The raw product was sieved and the size fraction between 100 and 140 μm was rinsed repeatedly with ultra-pure water to remove fine particles. The BET surface for this fraction was 0.20 m$^2$/g. Arsenic concentration in the calcite solid was below ICP detection limit. The calcite was stored in solution, in order to avoid artefacts due to exposure to air (Stipp et al., 1996). All the experiments were performed using the same calcite stock suspension and conducted at room temperature with 0.01 M background electrolyte medium. The pH was adjusted by adding either HNO$_3$ or NaOH.

### *2.1 Sorption Isotherms.*

Adsorption studies of As(III) were carried out at pH 7.5 by adding aliquots of As(III) stock solutions to calcite suspension in centrifugation tubes (40g/L). The suspensions with the As(III)-containing solution and blank solutions were placed in a reciprocal shaker at 25°C for 24 h. The pH of the suspensions and blank solutions was measured with a pH meter (Metrohm 781pH/ion Meter). After the reaction time the suspensions were centrifuged and filtered through 0.20-μm cellulose membranes afterwards. Reacted solid samples were washed with dionized water to remove the excess salts and air-dried, and stored until characterisation by X-ray diffraction (XRD).

*2.2 As-calcite co-precipitation.*

Calcite was precipitated at pH = 7.5 by addition of 0.5N $CaCl_2$ and 0.5N $Na_2CO_3$ solutions. Three samples were synthesised in presence of As(III), plus one pure calcite to be used as reference. The experiments were carried out at 20°C and the precipitates were aged for 2 h and, there after, washed 3 times with deionised water in order to eliminate NaCl. Solids were dried and prepared for X-ray and neutron diffraction analyses. Total chemical analyses of the solids were done after acid digestion using an ICP-AES.

*2.3 Chemical Analyses*

Inductively coupled plasma-optical emission spectroscopy (ICP-OES)(Perkin Elmer Optima 3000DV) was used to measure total arsenic concentrations (> 7.5 μM) in the solutions. Solutions with total arsenic concentrations lower than 7.5 μM were analysed by atomic fluorescence HPCL (Excalibur P S Analytical –HPCL Varian Pro Star).

*2.4 Diffraction Methods*

Neutron diffraction experiments were performed on a two-axis diffractometer (D20) located the High Flux Reactor of Institut Laue Langevin, Grenoble, France. A Cu200 monochromator was used obtaining a monochromatic neutron beam of 1.29095(7) Å, diffracted by samples and collected by a banana-like detector. The ³He multidetector consists of 1536 microstrip cells covering an angular range of 153.6º with a regular step of 0.1º. Samples were held in cylindrical Vanadium containers of 8.0 mm internal diameter and 0.2 mm wall thickness. Diffraction patterns were taken for the samples in their container and the empty cell, in the range of 10 to 130º. The precise value of neutron's wavelength and zero point of diffraction angle $2\theta_0 = 0.777(2)º$ were determined using an independent measurement of Si to calibrate.

X-ray diffraction data were acquired at ID11 beamline of ESRF, Grenoble, France. A heavy-duty high precision 8-circle diffractometer from Oxford Instruments was used. A Si (111) analyzer crystal coupled to a scintillation point detector was attached to the arm of the circle used. The gap width of the undulator was chosen to get a wavelength close to 0.52 Å, which after calibration it was found to be 0.51678 Å. The beam was collimated to get a rectangular section of 2 x 1.1 mm² at the centre of the circle. Powder specimens were introduced in Lindemann glass capillaries of 0.7 mm diameter. During the data acquisition the capillaries were spinned at 3000 rpm while the detector arm was displaced in an arc from 5 to 50º in scattering angle.

The measured neutron and X-ray diffraction spectra were processed with the full-pattern analysis Rietveld method, using the program FullProf (Rodríguez-Carvajal, 1997) for refining the crystal structure. We started by fitting the reference sample of pure calcite, for which we assumed a trigonal space group with standard atomic positions. The only fitted parameters were lattice constants (*a* and *c*), the *x* co-ordinate of O atoms and isotropic thermal constants for each atom. Our final results agree with reported values in the literature (Maslen *et al.*, 1993). We applied the same procedure to fit As-containing samples.

*2.5 Simulations*

Geometrical optimisations of the unit cell of pure calcite and of 2×2×1 and 3×2×1 supercells[1] of calcite doped with arsenic were done by using the Density Functional Theory (DFT) based code, Vienna Ab-initio Simulation Package (VASP), (Kresse and Furthmuller, 1999), (Kresse, 1996) in order to reproduce the expansion of the unit cell produced by the incorporation of As atoms into the structure of calcite. The Perdew-Burke-Enzerhof (PBE) variation of the General Gradient Approximation (GGA) to DFT was chosen to calculate the atomic ground state energy with VASP. This code uses a set of pseudo-potentials to represent the core electrons and a set of wave planes (whose cut-off energy can be set up) to represent the electrons from the outer shells. The free energy of the system is used as variational quantity. The program does several refinements of the electronic density before moving the ions according to Newton's laws.

The unit cell of pure calcite structure obtained from the Rietveld refinements of powder diffraction data was used as starting point for all the models. 2×2×1 and 3×2×1 supercells of calcite were doped with one and two atoms of arsenic, respectively, by replacing the carbon atoms sitting in the bulk of the supercells avoiding any surface effect in the volume expansion. The geometrical optimizations were made at zero temperature (to simulate the effects of temperature, VASP gives an initial speed to each atom accordingly to a Maxwellian distribution at the desired temperature; in this case, every atom had a zero initial speed). This fact led to a decreasing in volume of the unit cell of the optimized pure calcite of about a 1.5% of its experimental value. This effect on geometrical modeling has been already reported in the literature (Mihalkovic and Widom, 2004). The volume of the modeled unit cell of pure calcite was rescaled to its experimental value, obtained from simultaneous X-ray and neutron Rietveld refinements of diffraction data. All the models were rescaled by the same factor; however this effect does not affect our results as only relative volume changes are discussed.

---

[1] $\ell \times m \times n$ representing a supercell of $\ell$, $m$ and $n$ cells in directions *a*, *b* and *c*, respectively.

## 3. Results and discussion

### *3.1 Sorption-coprecipitation continuum isotherm*

In moderately oxidizing to moderately reducing environments, the aqueous species $H_3AsO_3^o$ or one of its deprotonated equivalents is considered as the dominant form of dissolved arsenic. Distribution of arsenite species as a function of pH at 25ºC and 1 bar is shown in Fig. 1. In the pH range of our experiment, 98% of As(III) is present as $As(OH)_3^o$. Above pH 9, As(III) behaves like a weak acid and its speciation (Fig.1) is the result of the following equilibria:

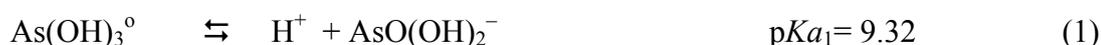

$$As(OH)_3^o \rightleftharpoons H^+ + AsO(OH)_2^- \qquad pKa_1 = 9.32 \qquad (1)$$

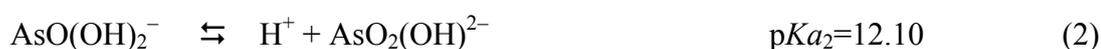

$$AsO(OH)_2^- \rightleftharpoons H^+ + AsO_2(OH)^{2-} \qquad pKa_2 = 12.10 \qquad (2)$$

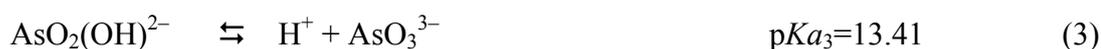

$$AsO_2(OH)^{2-} \rightleftharpoons H^+ + AsO_3^{3-} \qquad pKa_3 = 13.41 \qquad (3)$$

Experimental data of $\log \Gamma_{As(III)}$ *vs.* $\log [As(OH)_3^o/As_{sat}]$ are plotted in Fig.2, defining a curve whose low and high ends consist of asymptotic lines of slope 1 and which exhibits an inflexion point around $\Gamma_{As(III)} = 10^{-2}$. Following the approach of Farley et al. (1985), Comans and Middleburg (1987) and Wersin et al. (1989), the ordinate of the inflexion point is equal to $\log\left(\dfrac{S_T}{Tot\ CO_3}\right)$ which leads the surface site density ($S_T$). The site density value (6 sites/nm$^2$) is in agreement with previous sorption studies performed on calcite (Van Cappellen et al., 1993; Charlet, 1994; Pokrovski et al., 1996). Recent models have distinguished not one >CaCO$_3$ surface species, but two types of surface species (Van Cappellen et al., 1993; Pokrovski et al., 1996), namely >CaO$_3$H$^o$ and >CaOH$_2^+$ at equal total concentrations. The total surface site density of 9.8 sites/nm$^2$ used in these studies may indicate that both types of surface sites do participate to the sorption of $As(OH)_3^\circ$.

At pH 7.5, the neutral $As(OH)_3^o$ species predominates. Adsorption of this non-ionic species on a neutral, or slightly positively charged ionic solid is *a priori* not favored (Van Cappellen et al., 1993). From pH 7 to pH 9, the main chemical reaction is the dissolution of calcite (Plummer and Busenberg, 1982):

$$Ca^{2+} + H_2CO_{3(aq.)} = CaCO_3(s) + 2H^+ \quad\quad K_b \quad (4)$$

and the sorption of $H_3AsO_3^o$ on a $>CaCO_3$ neutral surface site may be written (Wersin et al., 1989):

$$H_3AsO_3^o + >CaCO_3 + Ca^{2+} = CaCO_3(s) + >CaHAsO_3 + 2H^+ \quad\quad K_c \quad (5)$$

Combination of the two reactions lead to:

$$K_v = \frac{K_c}{K_b} = \frac{[>CaHAsO_3]}{[>CaCO_3]} \frac{(H_2CO_3^o)}{(H_3AsO_3^o)} \quad\quad (6)$$

which describes the Vanselow-like exchange of neutral $H_3AsO_3^o$ and $H_2CO_3^o$ species at the surface of calcite. At As(III) concentrations lower than $10^{-5}$ M m$^{-2}$ the amount of sorbed As(III) was proportional to $As(OH)_3^o$ concentrations in solution. Assuming all sorbed As(III) to be present as $>CaHAsO_3^o$, the asymptotic line to the low concentration data corresponds to the Langmuir type adsorption as described by Eq. (5). The obtained values for $^cK_c$ and $K_b$ are equal to $1.5 \times 10^{-5}$ and $0.88 \times 10^{-6}$, respectively. The $K_v$ value is then evaluated to:

$$K_v = 1.88 \times 10^1 \quad\quad (7)$$

Therefore, the non deprotonated arsenious acid adsorption onto calcite is favored compared to that of the neutral carbonic acid species, which furthermore beyond pH 6.3 is a minor carbonate species (whereas $H_3AsO_3^o$ remains the dominant As(III) species up to pH 9). The derivation is available in Appendix A.

At pH 7.5, sorption still increases beyond a monolayer, *i.e.* for $\Gamma_{As(III)}$ larger than $10^{-2}$ (Fig. 2). In this range of concentrations the amount of $As(OH)_3$ adsorbed is proportional to the $As(OH)_3^o$ solution total concentration. This sorption beyond saturation of surface site but below the saturation with respect to any solid phase, is typical of an co-precipitation, and more specifically of an ideal solid solution (Farley et al., 1985). The end member of the ideal solid solution is formed according to:

$$Ca^{2+} + H_3AsO_3° = CaHAsO_3(s) + 2H^+ \qquad K_a \qquad (9)$$

A solubility product ($K_s$) for $CaHAsO_3$ was experimentally determined by Vandercasteele *et al.*, (2002) to be equal to $1.07 \times 10^{-7}$. Fit of the entire isotherm with the $\Gamma_{As(III)}$ equation derived in the appendix lead to a $^cK_a$ to be equal to $5.9 \times 10^{-9}$. The difference on $^cK_a$ value may be due to different experimental conditions in the two studies and to the fact that solid phase in the Vandercasteele *et al.*, (2002) study was not structurally identified as $CaHAsO_3$. In both cases, the high solubility of the calcium-arsenite this solid phase is clearly pointed out.

On the onset of precipitation ($\log\left(\dfrac{S_T}{TotCO_3}\right) > 0$), we observed the presence of a metastable phase, attributed to the fact that the composition of the surface phase can vary continuously from that of the pure substrate to that of the new phase (Stumm, 1992).

*3.3 As(III)-calcite co-precipitation*

In order to study specifically the As(III) incorporation into the calcite structure during co-precipitation, three co-precipitated samples (C1, C2, and C3, Fig. 3) were synthesized.

Our neutron and X-ray diffraction data show a small expansion of the unit cell as arsenic incorporation into calcite structure occurs (Fig. 3). Rietveld refinements of both X-ray and neutron diffraction data were carried out separately. The much smaller mass of sample used to get the X-ray diffraction pattern naturally gives a smaller surface-to-bulk ratio.

This fact together with the less penetrating nature of X-ray induces the emergence of diffraction peaks of phases that exist in very low proportion. In this case another crystalline phase of calcium carbonate (Vaterite) was observed in the X-ray diffraction patterns while the neutron patterns showed diffraction peaks corresponding to calcite solely. Neutron diffraction is a good complement to X-ray diffraction experiments since neutrons, being more penetrating, probes the bulk of a larger specimen, yielding diffraction data with better statistics. This is exactly what is needed to detect relatively scarce and randomly substituted As atoms present in the calcite crystal framework. Despite all that, our neutron and diffraction data, after being analyzed by Rietveld, agree in the cell parameters, giving similar cell volumes within the error bars fot both type of measurements (see Fig. 3).

The geometry of pure calcite was optimised, as well as that of supercells 2×2×1 (406 mM/kg and 791 (mM/kg) and 3×2×1 (273 mM/kg) models. In these supercells one and two $CO_3^{2-}$ ions were replaced by $HAsO_3^{2-}$ ions. The models showed an expansion in the volume proportional to the replacement of $CO_3^{2-}$ units by $HAsO_3^{2-}$ units, as expected by Vegard´s law (see Fig 4). This expansion is due to the increasing in the length of the lattice parameter *a* with the increasing concentration of arsenic. Also a displacement of 0.57 Å of the arsenic atom above the oxygen base along the [0001] direction was found (see Fig. 5) in agreement with previous reports (Cheng et al., 1999). These authors showed, by means of X-ray standing wave experiments, a displacement of the arsenic atom of 0.76 Å over the oxygen base along the [0001] direction at surface level. Worth noticing is the fact that neutron experiments and simulations give us access to bulk properties, showing how arsenic atoms accommodate not only at surface level but also in the whole structure of calcite.

The models give us a way to extrapolate the arsenic concentration in the bulk of the samples by comparing the relative variations of the volume of experimental and simulated data. The experimental values of the volume of unit cells were interpolated in the linear fit that describes the models' expansion of the volume (Fig. 4), giving values of 22 ± 6, 24 ±10 and 38±6 mM/kg of arsenic trapped in calcite structure for samples C1, C2 and C3, respectively (see Figure 3), or in average 30 ± 6 mM/kg of As(III) incorporated in calcite.

Simulations of 3×2×1 supercells with one $CO_3^{2-}$ unit replaced by one $HAsO_3^{-2}$ unit were done to check whether the replacement is more likely in sites on the same or in different crystallographic planes. The higher enthalpy of formation of $\Delta H = 211.36$ meV for the calcite structure with two As atoms lying on the same plane shows that replacement is more likely to happen in different planes as it leads to a more stable structure.

## 4. Conclusions

Macroscopic equilibrium data of As(III) sorption onto calcite depicts an S-shape sorption isotherm (defined here as $\log \Gamma_{As(III)}$ *vs.* $\log [As(OH)_3°]/As_{sat}]$. These data could be adequately described by a model originally derived by Farley et al., (1985) for metal oxides, which includes both adsorption and precipitation of an ideal solid solution as the two possible sorption mechanisms involved. The model was improved to be adapted to describe anion sorption and co-precipitation onto calcite. The inflexion point of the S-shaped isotherm was assumed to correspond to surface site saturation, and indicates that both types of calcite site surface site (>CaOH$^{+2}$ and >CaO$_3$H°) participate on As(OH)$_3$° sorption.

The formation of solid solution series between AsO$_3$HCa and CaCO$_3$ could not be fully evaluated because crystalline structure of AsO$_3$HCa is not known. However the obtained experimental results suggest that a solid solution precipitated at As(III) concentrations was higher than $1.5 \times 10^{-3}$ M. DFT model result of the atomic displacement of As atoms along [0001] direction are compatible with those published by Cheng et al., (1999). The slightly lower value of the displacement (0.57 Å *vs.* 0.76 Å) can be due to the fact that atoms near the surface are less attached to the solid and can be displaced with more freedom. However, the relative agreement confirms our model, which can be then used to evaluate the amount of As trapped in calcite at various arsenite concentrations.

Co-precipitation experiments show that As(III) is readily taken up to form solid solution with calcite. These results imply that the concentration of As(III) found in calcite in natural environments is limited by As(III) availability during crystal grown rather than by calcite ability to incorporate it into the bulk structure. The results also suggest that in a system where calcite is precipitating in presence of high As(III) concentrations, an average concentration of 30 ± 6 mM/kg of As(III) could be incorporated into calcite structure.

Acknowledgments

We thank Mark Johnson for his help and discussion on the numerical simulations, Gavin Vaughan for his assistance with the X-Ray diffraction experiments, and Delphine Tisserand for analytical determinations. This work was partially supported by the Action ECOS-SUD A02U01.

Figure Captions

Fig. 1: Arsenic (III) speciation. Taken from Wood et al., (2002).

Fig. 2: Sorption isotherm of $As^{3+}$ on calcite, ○= experimental data, solid line: Adsorption/precipitation model with $S_T$ = 6 sites/nm$^2$, $^cK_c$= 1.5x10$^{-5}$ , $^cK_a$= 5.9x10$^{-9}$ . Broken lines: lower (only adsorption) higher (only precipitation reactions). log $\Gamma_{(AsIII)}$ is defined as $\log\left(\frac{S_T}{Tot\,CO_3}\right)$ where $S_T$ is measured in M/l and (*Tot CO₃*) represents the carbonate site concentration under the experimental conditions of this study. As$_{sat}$ is defined as the As(III) concentration in equilibrium with CaHAsO$_{3\,(s)}$.

Fig. 3: Cell expansion with total As concentration in solids. Samples are labelled C0, C1, C2 and C3 (C0: pure calcite sample). Filled circles and empty triangles indicate X-ray diffraction and neutron diffraction data, respectively.

Fig. 4: Simulations of the expansion of 2×2×1 and 3×2×1 calcite supercells doped with one and two As atoms. The inset shows the interpolation of experimental data on the linear fit, which give us values of As concentration in the samples.

Fig. 5: Displacement of As atom along the [0001] direction, as seen in the simulations. The value of the displacement (0.57 Å) is smaller than that seen by Cheng et al., (1999) (0.76 Å) in X-ray standing wave experiments at the surface of calcite. Simulations are probing the bulk of the calcite structure and so the As atom has a more restricted pathway to displace.

Appendix A

Reaction A $\quad Ca^{2+} + H3AsO_3^\circ = CaHAsO_{3(s)} + 2H^+$

$$K_a = \frac{K_1}{\gamma_{CaHAsO_3}} = \frac{X_{CaHAsO_3}(H^+)^2}{(Ca^{2+})(H_3AsO_3^\circ)}$$

$$K_a = \frac{(H^+)^2}{(Ca^{2+})(H_3AsO_3^\circ)_{sat}}$$

$$z = X_{CaHAsO_3} = \frac{(H_3AsO_3^\circ)}{(H_3AsO_3^\circ)_{sat}}$$

with

$$(H_3AsO_3^\circ)_{sat} = \frac{(H^+)^2}{(Ca^{2+}) \cdot K_a}$$

Reaction B $\quad Ca^{2+} + H_2CO_3 = CaCO_{3(s)} + 2H^+$

$$K_b = \frac{K_2}{\gamma_{CaCO_3}} = \frac{X_{CaCO_{3(s)}}(H^+)^2}{(Ca^{2+})(H_2CO_3)} = \frac{(H^+)^2}{(Ca^{2+})(H_2CO_3)_{sat}}$$

$$H_2CO_3 = (H_2CO_3)_{sat} \, X_{CaCO_3}$$

$$= (H_2CO_3)_{sat} \, (1 - \frac{(H_3AsO_3)}{(H_3AsO_3)_{sat}})$$

with

$$(H_2CO_3)_{sat} = \frac{(H^+)^2}{K_b(Ca^{2+})}$$

$$(H_2CO_3) = \frac{X_{CaCO3}(H^+)^2}{K_b(Ca^{2+})} = (1-z)(H_2CO_3)_{sat}$$

Reaction C

$$H_3AsO_3 + >CaCO_3 + Ca^{2+} = CaCO_{3(s)} + >CaHAsO_3 + 2H^+$$

$$K_a = \frac{X_{CaHAs_3}(H^+)^2}{(Ca^{2+})(H_3AsO_3)}$$

$$K_b = \frac{X_{CaCO3}(H^+)^2}{(Ca^{2+})(H_2CO_3)}$$

$$K_c = \frac{X_{CaCO3}[>CaHAsO_3](H^+)^2}{(H_3AsO_3)[>CaCO_3](Ca^{2+})}$$

$$B = \frac{K_c}{K_a} = \frac{X_{CaCO_3}}{X_{CaHAsO_3}} \frac{[>CaHAsO_3]}{[>CaCO_3]}$$

$$K_v = \frac{K_c}{K_b} = \frac{[>CaHAsO_3]}{[>CaCO_3]} \frac{(H_2CO_3)}{(H_3AsO_3)}$$

$$S_T = [>CaHAsO_3] + [>CaCO_3]$$

$$= [>CaHAsO_3](1 + \frac{X_{CaCO_3}}{X_{CaHAsO_3}} B^{-1})$$

$$= [>CaHAsO_3]\left(1 + B^{-1}(\frac{1-z}{z})\right)$$

Thus:

$$[>CaHAsO_3] = S_T ((1-z)/(zB+1-z)$$

$$[>CaCO_3] = S_T (1 + (B^{-1}(z^{-1}-1))^{-1})^{-1}$$

$$[CaHAsO_{3(s)}] = [CaCO_{3(s)}] z/(1-z)$$

$$z = X_{CaHAsO_3} \frac{[CaHAsO_3]}{[CaCO_{3(s)}] + [CaHAsO_3]}$$

$$Tot\ CO_3 = [CaCO_{3(s)}] + [\rangle CaCO_3] + [H_2CO_3]$$

$$[CaCO_{3(s)}] = Tot\ CO_3 - S_T\ ((1-z)/(zB+1-z)) - (H_2CO_3^\circ)_{sat}\ (1-z)$$

so:

$$y = \Gamma = \frac{[\rangle CaHAsO_3]}{Tot\ CO_3} + \frac{[CaHAsO_{3(s)}]}{Tot\ CO_3}$$

is obtained by:

$$\frac{[\rangle CaHAsO_{3(s)}]}{TotCO_3} = \frac{S_T}{[TotCO_3]} \frac{Bz}{Bz+1-z}$$

$$\frac{[CaHAsO_{3(s)}]}{Tot\ CO_3} = \frac{[CaCO_3]}{[Tot\ CO_3]} \frac{z}{1-z}$$

$$\Gamma = \frac{z}{1-z}\left[1 - \frac{S_T}{Tot\ CO_3}\frac{(1-z)(1-\beta)}{zB+1-z} - \frac{H_2CO_3^{\circ\ sat}}{Tot\ CO_3}(1-z)\right]$$

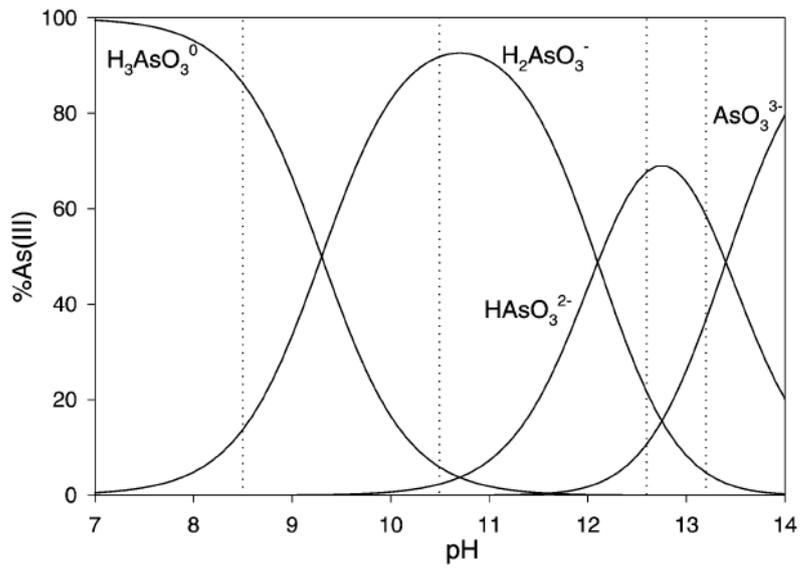

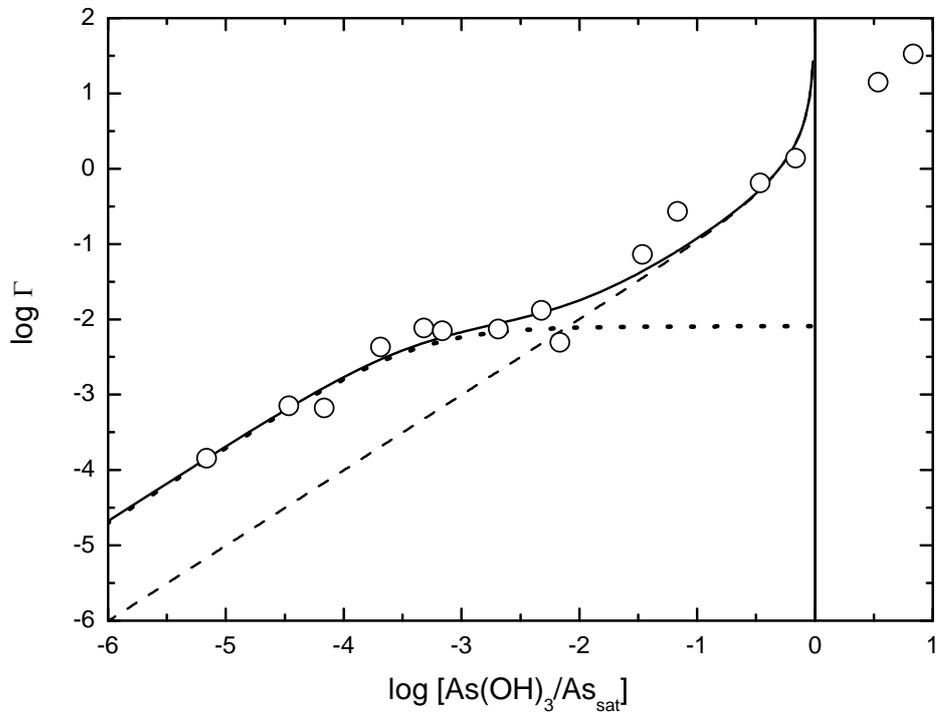

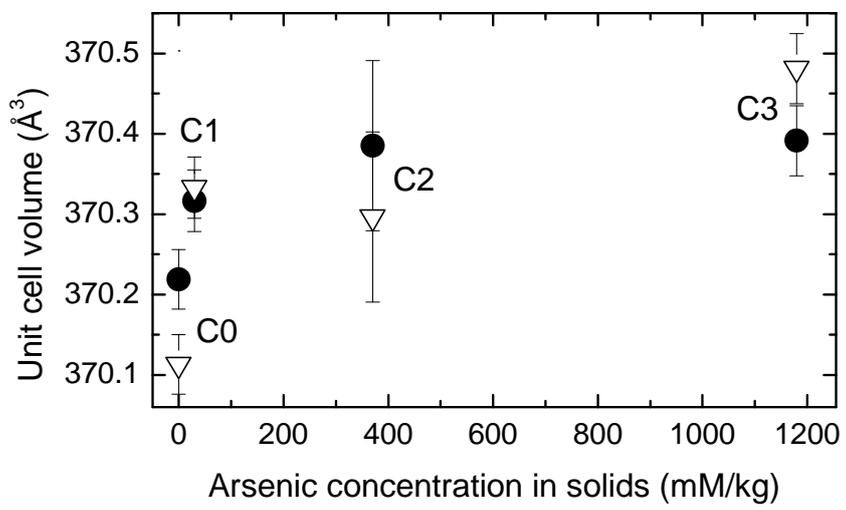

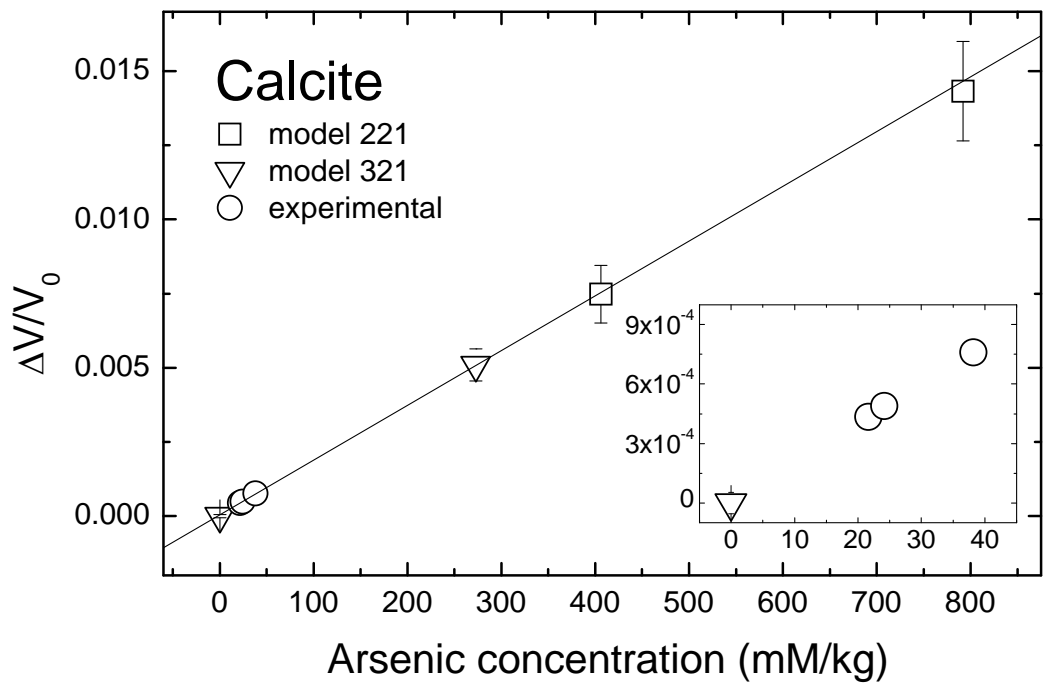

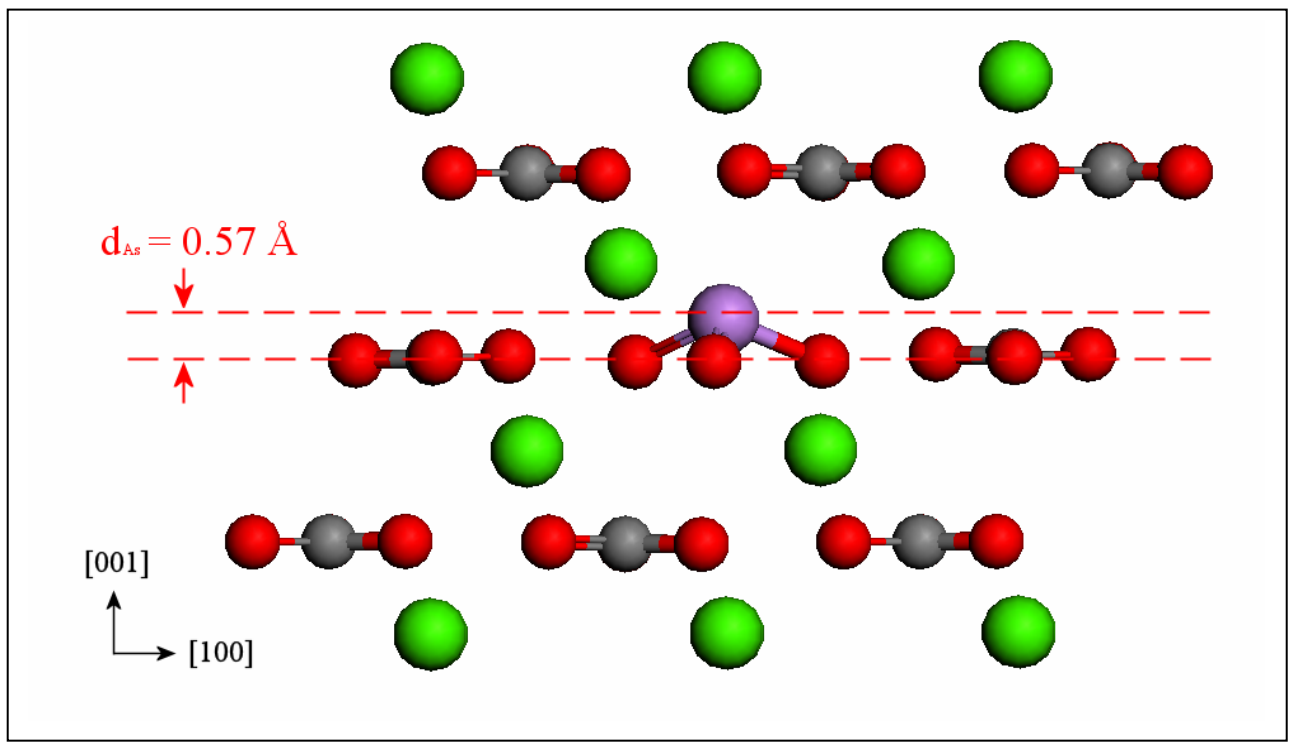